\documentstyle[11pt]{article}
\begin{document}

\begin{titlepage}
\begin{flushright}
{OUTP-98-09P}\\
{June 1998}\\
\end{flushright}
\vskip 0.5 cm
\begin{center}
 {\Large{\bf The effect of Yukawa couplings on Unification Predictions\\
\vskip 0.2cm
and the non-perturbative limit}}\\ 
\vskip 0.6 cm
  {{\large\bf G.~Amelino-Camelia$^{*,\dagger,}$\footnote{
                E-mail address: Giovanni.Amelino-Camelia@cern.ch},~ 
              Dumitru~Ghilencea$^{*,}$\footnote{
                E-mail address: D.Ghilencea1@physics.oxford.ac.uk
}}
and 
          {\bf Graham~G.~Ross$^{*,}$\footnote{
                E-mail address: G.Ross1@physics.oxford.ac.uk}}}
         \vskip 0.7 cm
\noindent
$^*$ {{\it Department of Physics, Theoretical Physics, University of Oxford}}\\
{\it 1 Keble Road, Oxford OX1 3NP, United Kingdom}\\
\medskip
$^\dagger$ {\it Institut de Physique, Universit\'e de Neuch\^atel,}\\
{\it rue Breguet 1, CH-2000 Neuch\^atel, Switzerland}\\
\end{center}
\vskip 2 cm
\begin{abstract}
We investigate the effects of Yukawa couplings on the phenomenological
predictions for a class of supersymmetric models which 
allows for  the presence  of complete  $SU(5)$ multiplets in addition to the
Minimal Supersymmetric Standard Model spectrum.
We develop a two-loop analytical approach to quantify the predictions for gauge unification including Yukawa couplings.  The effects of
the heavy thresholds of the model are also included.
In some cases accurate predictions can be made for the unification scale, 
irrespective of the initial (unknown) Yukawa couplings, so long as 
perturbation theory remains valid. We also consider the limit of a large
number of extra states and compute the  predictions in a resummed
perturbation series approach to show that the results are stable in
this limit. Finally we consider the possibility of making
predictions for the case the gauge and Yukawa couplings enter the 
non-perturbative domain below the unification scale 
and estimate the errors which affect these predictions.
\end{abstract}
\end{titlepage}

\setcounter{footnote}{0}

\section{Introduction}

In a recent paper \cite{grl} the phenomenological implications of a class of
string motivated supersymmetric models based on level-1 heterotic string
models with Wilson line breaking of the underlying E$_{6}$ symmetry were
investigated. These models (henceforth referred to as Extended Minimal
Supersymmetric Models (EMSSM)) contain additional vector-like
representations filling out complete five and ten dimensional
representations of $SU(5)$ even though the gauge group is just that of the
Standard Model\footnote{%
To be precise these models contain additional vector-like states $I+\bar{I}$
, where $I$, $\bar{I}$ denote complete representations of $SU(5)$. The
components $\psi $ of the additional complete $SU(5)$ representations
transform under $SU(3)$ and $SU(2)$ groups as follows: for $\psi =d^{c}:(%
\bar{3},1)\ $, $\psi =l:(1,2)$ for the case $I$ is the 5 dimensional
representation of $SU(5)$ and $\psi =e^{c}:(1,1)\ ,\psi =u^{c}:(\bar{3},1)\
,\psi =q:(3,2)\ $ for the case $I$ is the 10 dimensional representation of $%
SU(5)$}. This class of string models has several advantages for model
building \cite{witold1};\ the Wilson line breaking necessary to break the $%
E_{6}$ symmetry in level-1 theories offers an elegant way out of the
doublet-triplet splitting problem encountered in GUTs. It can also lead to
the standard unification of the gauge couplings without the need of a Grand
Unified Group below the compactification scale\footnote{%
Alternative constructions were developed which can lead to non-standard U(1)
normalisation even in level-1 theory - see discussion in ref. \cite{dienes1}}%
. In this case the string prediction for the scale of unification of the
gauge couplings may be directly compared with the value obtained continuing
the gauge couplings up in energy providing the possibility of a quantitative
test of string unification including gravity.

The effect of the additional vector-like states on $\alpha _{3}(M_{z})$ and
the value of the unification scale was explored in detail in\cite{grl}. It
was found that, working to two-loop order in the gauge sector, the
unification scale is systematically increased by a small factor (about 3 or
less, a function of the value of the unified coupling and of the number $%
n=(n_{5}+3n_{10})/2$ of pairs \footnote{$n_{5}=N_{5}+N_{\overline{5}}$ and $%
n_{10}=N_{10}+N_{\overline{10}}$} of additional complete $SU(5)$
representations), while the strong coupling was systematically increased,
taking it further from the experimental value \cite{particledata}, $\alpha
_{3}(M_{z})=0.118\pm 0.003$. The increase in the unification scale does not,
in fact, take it closer to the weakly coupled heterotic string prediction 
\cite{scale} because the increase is correlated with an increase in the
unified coupling which always happens when additional matter is added and
the heterotic string prediction also increases with the unified coupling. As
a result there is still a discrepancy\footnote{Ways to reconcile 
this scale discrepancy have been reviewed by Dienes \cite{dienes2}.}
 of a factor of 10-20. 
However, in this analysis the effects of Yukawa couplings on
the running of the gauge couplings were not included and we seek to include
them in this paper. An immediate difficulty arises because in theories with
additional massive vectorlike states there may be many new Yukawa couplings
involving these heavy states. We shall consider two basic possibilities
which give an indication of the general possibilities and uncertainties. In
the first we include only the Yukawa couplings present in the MSSM\ which
are responsible for the third generation masses. In the second we include
couplings between massive Higgs doublets and the light quarks and leptons.
This model goes some way towards explaining the light quark mass matrix via
mixing in the Higgs sector \cite{ibanezross} as well as  explaining the
masses of the third generation, in a ``fixed-point-scenario'' \cite{sun}. We
also comment on how our results would be affected if further couplings
involving massive states were included.

The paper is organized as follows. In Section 2 we review the effect of
heavy thresholds on the running coupling constants and we analyze the
renormalisation group evolution (RGE) predictions for the gauge couplings
with Yukawa couplings effects included. In section 3 we derive an analytic
form for $\alpha _{3}(M_{z})$ and the unification scale as well as the
decoupling scale of the extra matter and make some numerical estimates. In
section 4 we consider the case of large $n.$ Finally, in Section 5 we
consider the case of unification at strong coupling. We discuss the
differences between the (quasi)-fixed point approach \cite{sun} and the two
loop perturbative approach results. Our conclusions are presented in the
last section.

\section{Heavy thresholds' contribution from the integrated NSVZ beta
function}

In this section we make use of the Novikov-Shifman-Vainshtein-Zakharov
(NSVZ) beta function \cite{NSVZ,murayama2} for the gauge couplings, which
was derived using holomorphicity arguments and the instanton calculus. We
integrate it in the presence of the additional heavy states of the EMSSM\
model and then use its two-loop approximation to perform further
calculations.

The integrated expression for the gauge couplings' running, in the absence
of these heavy thresholds due to the presence of the extra heavy states we
consider, was presented in ref. \cite{shifman} and was deduced without
actually integrating the beta function, but by using physical arguments and
the properties of the holomorphic gauge coupling; from that, the integrated
form in the presence of our additional heavy thresholds can be easily
``guessed''. For clarity, we prefer to re-derive it here by direct
integration. For a fuller discussion the reader is referred to the
literature \cite{murayama2,shifman,murayama1}. We have \cite{NSVZ} 
\begin{equation}
\beta (\alpha )^{NSVZ}\equiv \frac{d\alpha }{d(\ln \mu )}=-\frac{\alpha ^{2}%
}{2\pi }\left[ 3T(G)-\sum_{\sigma }^{{}}T(R_{\sigma })(1-\gamma _{\sigma
}^{NSVZ})\right] \left( {1-T(G)\frac{\alpha }{2\pi }}\right) ^{-1}
\label{shifmanbetaalpha}
\end{equation}
with the definition ($\mu $ is the running scale) 
\[
\gamma _{\sigma }^{NSVZ}=-\frac{d\ln Z_{\sigma }}{d\ln \mu }
\]
and where $T(G)$ and $T(R_{\sigma })$ represent the Dynkin index for the
adjoint representation and for $R_{\sigma }$ representation respectively
(not necessarily the fundamental one), whose values are given in the
Appendix. The above sum runs over {\it all} matter fields $\sigma $ in
representation $R_{\sigma }$ and this includes, for example, the extra heavy
states (which we called $\psi $ -see footnote (1)), in addition to the low
energy spectrum of MSSM. This gives 
\begin{equation}
2\pi \frac{d\alpha ^{-1}}{d\ln \mu }+T(G)\frac{d\ln \alpha }{d\ln \mu }%
=3T(G)-\sum_{\sigma }T(R_{\sigma })(1-\gamma _{\sigma }^{NSVZ})
\end{equation}
which can be integrated from the high scale $M$ down to the low scale $\mu $
with $\mu >\Lambda _{Susy}$ to give 
\begin{equation}
2\pi (\alpha ^{-1}(M)-\alpha ^{-1}(\mu ))+T(G)\ln \frac{\alpha (M)}{\alpha
(\mu )}=3T(G)\ln \frac{M}{\mu }-\sum_{\sigma }\int_{\mu }^{M}T(R_{\sigma
})(1-\gamma _{\sigma }^{NSVZ})d(\ln {\tilde{\mu}})  \label{part}
\end{equation}
In the above equation $\alpha $ stands for any gauge group coupling. Any
extra heavy state $\psi $, with mass $\mu _{\psi }(\mu _{\psi })$ larger
than $\mu $, decouples at its {\it physical mass scale} in the running of
the {\it canonical} gauge coupling \cite{murayama2,murayama1} and therefore,
below $\mu _{\psi }$ scale, $\gamma _{\psi }=0$. 
From eq.(\ref{part}) we obtain the result 
\begin{eqnarray}
\alpha ^{-1}(\mu )=\alpha ^{-1}(M) &+&\frac{-3T(G)}{2\pi }\ln \frac{M}{\mu
\left( \frac{\alpha (M)}{\alpha (\mu )}\right) ^{1/3}}  \nonumber \\
&+&\frac{1}{2\pi }\sum_{\phi }^{{}}T(R_{\phi })\ln \frac{M}{\mu }+\frac{1}{%
2\pi }\sum_{\phi }^{{}}T(R_{\phi })\ln \frac{Z_{\phi }(M)}{Z_{\phi }(\mu )} 
\nonumber \\
&+&\frac{1}{2\pi }\sum_{\psi }^{{}}T(R_{\psi })\ln \frac{M}{\mu _{\psi }}+%
\frac{1}{2\pi }\sum_{\psi }^{{}}T(R_{\psi })\ln \frac{Z_{\psi }(M)}{Z_{\psi
}(\mu _{\psi })}
\end{eqnarray}
where $\phi $'s stand for a MSSM-like spectrum. At this point we make the
observation that 
\[
\mu _{\psi }Z_{\psi }(\mu _{\psi })=\mu _{\psi }^{o}
\]
which is the mass renormalisation equation, where $\mu _{\psi }^{o}$
represents the bare mass of the $SU(5)$ component $\psi $. For the case of a
Grand Unified Group Model, the gauge invariance principle requires the mass
terms $\mu _{\psi }\psi \overline{\psi }$ be invariant, which is possible
only if all bare masses are equal to a common value, $\mu _{g}$. We also
consider, without any loss of generality that $Z_{\psi }(M_{g})=Z_{\phi
}(M_{g})=1$ where $M_{g}$ is the value of unification scale. Taking $M=M_{g}$%
, we find 
\begin{eqnarray}
\alpha ^{-1}(\mu ) &=&\alpha ^{-1}(M_{g})+\frac{-3T(G)}{2\pi }\ln \frac{M_{g}%
}{\mu \left( \frac{\alpha (M_{g})}{\alpha (\mu )}\right) ^{1/3}}+\sum_{\phi
}^{{}}\frac{T(R_{\phi })}{2\pi }\ln \frac{M_{g}}{\mu Z_{\phi }(\mu )}+\frac{n%
}{2\pi }\ln \frac{M_{g}}{\mu _{g}}  \label{integrated} \\
&&+\frac{T(R_{\Sigma })}{2\pi }\ln \frac{M_{g}}{\mu _{\Sigma }}  \nonumber
\end{eqnarray}
with $n=(n_{5}+3n_{10})/2$,$\;n_{5}=N_{5}+N_{\overline{5}}$ and $%
n_{10}=N_{10}+N_{\overline{1}0}$. The last term in eq.(\ref{integrated})
could stand for the Higgs in the adjoint representation or for the $SU(3)$
Higgs 
triplet\footnote{These exotic Higgs effects 
will not be considered further in this paper}.
This formula is valid to all orders in perturbation theory, as long as
Supersymmetry is not broken. This equation was used, in this form, as a
check of the intermediate results obtained in \cite{grl}. 

In two-loop order for gauge couplings running,
$Z$ factors depend on the mean mass of the extra
heavy states which can be considered to be their common bare mass $\mu_g$
since the difference would be of three loop order. Hence, in two-loop
order  $\mu _{g}$ is the only mass scale associated with the 
heavy states in eq.(\ref
{integrated}). Therefore one sees that  the decoupling of 
the heavy states happens, in two-loop order,
 at $\mu _{g}$ and not at each of the physical masses of the
component fields of the massive $SU(5)$ representations. 

\section{Analytical Results from Renormalisation Group Evolution}

In this section we consider the effects of the Yukawa couplings on the
running of the gauge couplings and their effects on the predictions for the
unification scale, the strong coupling at electroweak scale and the scale
where the heavy states decouple. In the case the Yukawa couplings are
negligible, we recover previous two-loop analytical results \cite{grl}.

To keep our approach general and to exhibit the effects Yukawa couplings
have on the running of the gauge couplings, it is convenient to use eq.(\ref
{integrated}) for the running of the gauge couplings. We should mention that
in two-loop order this formula has no regularization ambiguities, which only
arise at three loop and beyond \cite{jones}. The advantage of eq.(\ref
{integrated}) for the evolution of the couplings is that it expresses the
running of the gauge couplings in terms of the wavefunction renormalisation
coefficients $Z_{\phi }$. Unlike the gauge wave function contribution which
is one-loop exact, one still needs to use perturbation theory to compute the 
$Z_{\phi }$ order by order for the matter wave function. However, for a two
loop approximation for the gauge couplings, only a one-loop calculation of $%
Z_{\phi }$ is required, simplifying the calculation. Their evolution from
the unification scale where they are normalised to unity\footnote{$Z_{\phi }$
depend on the unification scale!}, to the decoupling scale $\mu _{g}$ of the
heavy states is affected by the presence of $n=(n_{5}+3n_{10})/2$ {\it pairs}
of complete $SU(5)$ multiplets which affects the anomalous dimensions of the
matter fields\footnote{%
The running of the gauge couplings depends on $n_{5}+3n_{10}$ only and not
separately on $n_{5}$ or $n_{10}$ \cite{grl}.}.

Below the supersymmetry breaking scale we have to include the effect of the
low energy supersymmetric thresholds. Their effect at one-loop, which
corresponds to a two-loop effect if the splitting of the super-partner
masses is of radiative origin, we denote by $\delta _{i}$. This term is
needed because eq.(\ref{integrated}) is valid only as long as supersymmetry
is unbroken \footnote{$\delta _{i}$ include regularisation scheme conversion
factors as well.}.

With these considerations we rewrite eq. (\ref{integrated}) for the three
gauge couplings (j is a generation index) in the following way: 
\begin{eqnarray}  \label{integratedalpha}
\alpha _{i}^{-1}(M_{z}) &=&-\delta _{i}+\alpha _{g}^{-1}+\frac{b_{i}}{2\pi }
\ln \frac{M_{g}}{M_{z}}+\frac{n}{2\pi }\ln \frac{M_{g}}{\mu _{g}}-\frac{
\beta _{i,H_{1}}}{2\pi }\ln Z_{H_{1}}(M_{z})-\frac{\beta _{i,H_{2}}}{2\pi }
\ln Z_{H_{2}}(M_{z})  \nonumber \\
&&-\frac{\beta _{i,g}}{2\pi }\ln \left[ \frac{\alpha _{g}}{\alpha _{i}(M_{z})%
}\right] ^{1/3}-\sum_{j=1}^{3}\sum_{\phi _{j}}{}\frac{\beta _{i,\phi _{j}}}{%
2\pi }\ln Z_{\phi _{j}}(M_{z})
\end{eqnarray}
where $b_{1}=33/5$, $\,b_{2}=1$, $\,b_{3}=-3$ and where $\beta _{i,\phi
_{j}}\equiv T(R_{\phi _{j}}^{i})$, $i=\{1,2,3\}$, are the contributions to
one-loop beta function\footnote{%
We also used that the one-loop beta function is $b=-3 T(G)+ \sum T(R_\phi)$,
where the sum runs over all chiral supermultiplets in representation $R_\phi$%
.} of the matter fields $\phi _{j}$ (j=generation index), while $\beta
_{i,g}\equiv T^{i}(G)$ is the one-loop beta function for the pure gauge
(+gaugino) sector; the Higgs (+higgsino) sector contribution is included
separately via the terms proportional to $\beta _{i,H1,2}$ considered. Thus
we have

\begin{equation}
\beta _{i,\phi _{j}}=\left( 
\begin{array}{ccccc}
\frac{3}{10} & \frac{1}{10} & \frac{3}{5} & \frac{4}{5} & \frac{1}{5} \\ 
&  &  &  &  \\ 
\frac{1}{2} & \frac{3}{2} & 0 & 0 & 0 \\ 
&  &  &  &  \\ 
\ 0 & 1 & 0 & \frac{1}{2} & \frac{1}{2}
\end{array}
\right) _{i,\phi _{j}}\;\;\;\;\;\;\;\beta _{i,g}=\left( 
\begin{array}{c}
0 \\ 
\\ 
-6 \\ 
\\ 
-9
\end{array}
\right) ;\;\;\;\;\;\;\;\beta _{i,H_{1,2}}=\left( 
\begin{array}{c}
\frac{3}{10} \\ 
\\ 
\frac{1}{2} \\ 
\\ 
0
\end{array}
\right)
\end{equation}
independent of the values of $j$. The field $\phi _{j}$ runs over the set $%
\phi _{j}=\{l_{L},q_{L},e_{R},u_{R},d_{R}\}_{j}$, in this order, with $j$ as
generation index.

\subsection{$\alpha _{3}(M_{Z}),$ $M_{g}$ and $\mu _{g}.$}

To compute the two-loop values for the strong coupling
at electroweak scale, the unification scale and 
the decoupling scale  $\mu _{g}$ of the
extra massive states\footnote{$\mu_g$ is the decoupling scale only in two
loop order}, we use the two step method developed in reference \cite{grl}.

The first step eliminates the low energy supersymmetric thresholds
dependence by expressing our results as a change to MSSM predictions for
the strong coupling and the unification scale; this is possible because, to two
loop order for the RGE, the $\delta _{i}$'s are the same in both cases. Note
that this means that our results will be expressed in terms of $\alpha
_{g}^{o}$, $M_{g}^{o}$ and $\alpha _{i}^{o}(M_{z})$ which {\it all} have the 
$\delta _{i}$ dependence {\it included}.

The second step is more technical and consists of replacing the arguments of
the two-loop log's by the one-loop approximation. This will, of course,
generate the evolution of the couplings correct to two-loop order.

To implement the first step note that in the MSSM we have 
\begin{eqnarray}
\alpha _{i}^{o-1}(M_{z}) &=&-\delta _{i}+\alpha _{g}^{o-1}+\frac{b_{i}}{2\pi 
}\ln \frac{M_{g}^{o}}{M_{z}}-\frac{\beta _{i,H_{1}}}{2\pi }\ln
Z_{H_{1}}^{o}(M_{z})-\frac{\beta _{i,H_{2}}}{2\pi }\ln Z_{H_{2}}^{o}(M_{z}) 
\nonumber \\
&&-\frac{\beta _{i,g}}{2\pi }\ln \left[ \frac{\alpha _{g}^{o}}{\alpha
_{i}^{o}(M_{z})}\right] ^{1/3}-\sum_{j=1}^{3}\sum_{\phi _{j}}{}\frac{\beta
_{i,\phi _{j}}}{2\pi }\ln Z_{\phi _{j}}^{o}(M_{z})  \label{mssm}
\end{eqnarray}
where we employed the index ``o'' to label all MSSM related quantities, to
distinguish them from those of the extended model which introduces
additional heavy states. Note that $Z^{o}$ are normalised to unity at the $%
M_{g}^{o}$, $Z^{o}(M_{g}^{o})\equiv Z^{o}(0)=1$.

As is commonly done in all ``bottom-up'' approaches, we impose the
conditions $\alpha _{1}(M_{z})=\alpha _{1}^{o}(M_{z})$ and $\alpha
_{2}(M_{z})=\alpha _{2}^{o}(M_{z})$ and equal to their experimental values.
We now determine the change of 
 the strong coupling at electroweak scale; from (\ref
{integratedalpha}), (\ref{mssm}) we get that (i=1,2): 
\begin{eqnarray}
0 &=&\alpha _{g}^{-1}-\alpha _{g}^{o-1}+\frac{b_{i}}{2\pi }\ln \frac{M_{g}}{%
M_{g}^{o}}+\frac{n}{2\pi }\ln \frac{M_{g}}{\mu _{g}}-\frac{\beta _{i,H_{1}}}{%
2\pi }\ln \left[ \frac{Z_{H_{1}}(M_{z})}{Z_{H_{1}}^{o}(M_{z})}\right] -\frac{%
\beta _{i,H_{2}}}{2\pi }\ln \left[ \frac{Z_{H_{1}}(M_{z})}{%
Z_{H_{1}}^{o}(M_{z})}\right]  \nonumber \\
&&-\frac{\beta _{i,g}}{2\pi }\ln \left[ \frac{\alpha _{g}}{\alpha _{g}^{o}}%
\right] ^{1/3}-\sum_{j=1}^{3}\sum_{\phi _{j}}{}\frac{\beta _{i,\phi _{j}}}{%
2\pi }\ln \left[ \frac{Z_{\phi _{j}}(M_{z})}{Z_{\phi _{j}}^{o}(M_{z})}\right]
\label{dif12} \\
&&  \nonumber \\
\delta \alpha _{3}^{-1}(M_{z}) &=&\alpha _{g}^{-1}-\alpha _{g}^{o-1}+\frac{%
b_{3}}{2\pi }\ln \frac{M_{g}}{M_{g}^{o}}+\frac{n}{2\pi }\ln \frac{M_{g}}{\mu
_{g}}-\frac{\beta _{3,H_{1}}}{2\pi }\ln \left[ \frac{Z_{H_{1}}(M_{z})}{%
Z_{H_{1}}^{o}(M_{z})}\right] -\frac{\beta _{3,H_{2}}}{2\pi }\ln \left[ \frac{%
Z_{H_{2}}(M_{z})}{Z_{H_{2}}^{o}(M_{z})}\right]  \nonumber \\
&&-\frac{\beta _{3,g}}{2\pi }\ln \left[ \frac{\alpha _{g}\alpha
_{3}^{o}(M_{z})}{\alpha _{g}^{o}\alpha _{3}(M_{z})}\right]
^{1/3}-\sum_{j=1}^{3}\sum_{\phi _{j}}{}\frac{\beta _{3,\phi _{j}}}{2\pi }\ln %
\left[ \frac{Z_{\phi _{j}}(M_{z})}{Z_{\phi _{j}}^{o}(M_{z})}\right]
\label{al3}
\end{eqnarray}
where by $\delta \alpha _{3}^{-1}(M_{z})$ we denoted the (two-loop induced)
difference $\delta \alpha _{3}^{-1}(M_{z})=1/\alpha _{3}(M_{z})-1/\alpha
_{3}^{o}(M_{z})$. We note that the term $\ln(\alpha^o_3(M_z)/\alpha_3(M_z)$
can simply be neglected in the two-loop approximation as $\alpha^o_3(M_z)$
and $\alpha_3(M_z)$ are equal in one-loop, and therefore this term would
bring a higher order correction. We solve the system of  
eqs. (\ref{dif12}), (\ref{al3}) 
for $\delta \alpha _{3}^{-1}(M_{z})$, $M_{g}$ and $\mu _{g}$ in
function of $\alpha _{g}$ to get:
\begin{equation}
\delta \alpha _{3}^{-1}(M_{z})=-\frac{\sigma _{g}}{2\pi }\ln \left[ \frac{%
\alpha _{g}}{\alpha _{g}^{o}}\right] ^{\frac{1}{3}}-\sum_{j=1}^{3}\sum_{\phi
_{j}}\frac{\sigma _{\phi _{j}}}{2\pi }\ln \left[ \frac{Z_{\phi _{j}}(M_{z})}{%
Z_{\phi _{j}}^{o}(M_{z})}\right] -\frac{\sigma _{H_{1}}}{2\pi }\ln \left[ 
\frac{Z_{H_{1}}(M_{z})}{Z_{H_{1}}^{o}(M_{z})}\right] -\frac{\sigma _{H_{2}}}{%
2\pi }\ln \left[ \frac{Z_{H_{2}}(M_{z})}{Z_{H_{2}}^{o}(M_{z})}\right]
\label{mat1} \\
\end{equation}
where $\sigma _{\phi _{j}}=\beta _{1,\phi
_{j}}(b_{2}-b_{3})/(b_{1}-b_{2})+\beta _{2,\phi
_{j}}(b_{3}-b_{1})/(b_{1}-b_{2})+\beta _{3,\phi _{j}}$ and we have a similar
definition for $\sigma _{H_{1}}$, $\sigma _{H_{2}}$ and $\sigma _{g}$. Thus
\begin{equation}
\sigma =\left\{ l_{L_{j}}:\frac{-9}{14},\;\;q_{L_{j}}:\frac{-3}{2},\;\;H_{1}:%
\frac{-9}{14},\;H_{2}:\frac{-9}{14},\;\;e_{R_{j}}:\frac{3}{7},\;\;u_{R_{j}}:%
\frac{15}{14},\;\;d_{R_{j}}:\frac{9}{14},\;\;g:\frac{9}{7}\right\}
\end{equation}
We observe that $\sigma _{\phi _{j}}$ is negative for $SU(2)$ doublets and
positive for $SU(2)$ singlets and these will therefore have opposite effects
on the values of $\alpha _{3}(M_{z})$, for same sign of the associated
logarithmic factor. Moreover, the contribution to the strong coupling
 depends on
whether $Z$'s are larger or smaller than unity, which is determined by the
relative effect of gauge and Yukawa contributions to their running. This
should become clearer later. The overall effect on the strong coupling depends
however on the {\it relative} magnitude of the wavefunction
coefficients, $Z, $ of the extended model compared to those of 
MSSM, $Z^{o}$.

Using the notation $\Delta \beta _{12,\phi _{j}}=\beta _{1,\phi _{j}}-\beta
_{2,\phi _{j}}$ and $\Delta \beta _{12,H_{1,2}}=\beta _{1,H_{1,2}}-\beta
_{2,H_{1,2}}$ the change in the unification scale factor is given by (see
eqs.(\ref{dif12}), (\ref{al3})): 
\begin{equation}
\ln \frac{M_{g}}{M_{g}^{o}}=\frac{15}{14}\ln \left[ \frac{\alpha _{g}}{%
\alpha _{g}^{o}}\right] ^{\frac{1}{3}}+\frac{5}{28}\sum_{j=1}^{3}\sum_{\phi
_{j}}^{{}}\Delta \beta _{12,\phi _{j}}\ln \left[ \frac{Z_{\phi _{j}}(M_{z})}{%
Z_{\phi _{j}}^{o}(M_{z})}\right] +\left\{ \frac{5}{28}\Delta \beta
_{12,H_{1}}\ln \left[ \frac{Z_{H_{1}}(M_{z})}{Z_{H_{1}}^{o}(M_{z})}\right]
+H_{1}\leftrightarrow H_{2}\right\}  \label{mat2}
\end{equation}
Note that $\Delta \beta _{12,\phi _{j}}$ has negative sign for $SU(2)$
doublets and positive for $SU(2)$ singlets (just like $\sigma _{\phi _{j}})$
and hence, for same sign of the logarithmic factor, these terms will drive
the unification scale in opposite directions. Moreover, comparing eqs.(\ref
{mat1}) and (\ref{mat2}), we observe that the unification scale $M_g$ is
increased for positive $\Delta\beta_{12}$ eq.(\ref{mat2}), while alpha
strong eq.(\ref{mat1}) is also increased\footnote{$\Delta\beta_{12,\phi_j}$
has same sign as $\sigma_{\phi_j}$} for a given (positive) 
sign of the log's. Thus, we
already see it is difficult to decrease the strong coupling and increase
unification scale {\it simultaneously}. One could argue that $Z$ factors
have an implicit unification scale dependence as well, and that our above
explanation does not apply; however this dependence comes in under the log,
and is therefore very mild. Note that this applies independent of the type
of Yukawa interaction, implicitly present in the values of $Z$ coefficients.
This seems to be generic to models which consider {\it complete} additional
representations and it was also explored in \cite{grl} in the absence of
Yukawa effects with same conclusions. The above discussion should become
clearer later when a more quantitative study will be made.

Finally, the decoupling scale $\mu _{g}$ is given by (see eqs.(\ref{dif12}),
(\ref{al3})): 
\begin{equation}
\ln \frac{\mu _{g}}{M_{g}^{o}}=\frac{2\pi }{n}\left[ \frac{1}{\alpha _{g}}-%
\frac{1}{\alpha _{g}^{o}}\right] +\frac{\Omega _{g}}{n}\ln \left[ \frac{%
\alpha _{g}}{\alpha _{g}^{o}}\right] ^{\frac{1}{3}}+\sum_{j=1}^{3}\sum_{\phi
_{j}}^{{}}\frac{\Omega _{\phi _{j}}}{n}\ln \left[ \frac{Z_{\phi _{j}}(M_{z})%
}{Z_{\phi _{j}}^{o}(M_{z})}\right] +\left\{ \frac{\Omega _{H_{1}}}{n}\ln %
\left[ \frac{Z_{H_{1}}(M_{z})}{Z_{H_{1}}^{o}(M_{z})}\right]
+H_{1}\leftrightarrow H_{2}\right\}  \label{mat3}
\end{equation}
where 
\begin{equation}
\Omega _{F}=\frac{b_{2}^{\prime }\beta _{1,{F}}-b_{1}^{\prime }\beta _{2,{F}}%
}{b_{1}-b_{2}};\;\;\;\;\;\Omega _{g}=\frac{b_{2}^{\prime }\beta _{1,{g}%
}-b_{1}^{\prime }\beta _{2,{g}}}{b_{1}-b_{2}};
\end{equation}
with $F\equiv \{\phi _{j},H_{1},H_{2}\}$ and $b_{i}^{\prime }=b_{i}+n$.

To simplify our expressions further we need to compute the wavefunction
renormalisation coefficients. Above the mass of the additional states in the
extended model the running of these coefficients is altered implying that
the anomalous dimensions of the matter fields are changed. We have (with $%
t=1/(2\pi )\ln (scale)$) 
\begin{equation}
\frac{d}{dt}\ln Z_{F}(t)\equiv -4\pi \gamma_{F}=2\sum_{k=1}^{3}C_{k}({F})
\alpha _{k}(t)+\sum_{\nu =\tau ,b,t}^{{}}{\tilde{{\cal A}}}_{\nu }({F}%
)y_{\nu }(t) +{\cal O}(\alpha ^{2})  \label{Zs}
\end{equation}
In the above equation $C_{j}({F})$ is the second order Casimir operator for
the field ${F}$ for the group $SU(j)$ and is generation independent. The
coefficients ${\tilde{{\cal A}}}_{\nu }({F})$ depend on the type of the
superpotential of the model. For various cases we consider, their values are
given in the Appendix.

The above equation can easily be integrated to give full analytical
expressions for the wavefunction coefficients, valid in one-loop
approximation which is consistent with a two-loop running for the gauge
couplings. The solution is of the type: 
\begin{equation}
Z_{F}(t)=Z_{F}^{G}(t)\times Z_{F}^{Y}(t)  \label{solutie}
\end{equation}
where $Z_{F}^{G}$ and $Z_{F}^{Y}$ are given by the gauge and Yukawa running
parts respectively, with $Z_{F}^{G}(0)=1$ and $Z_{F}^{Y}(0)=1$. $Z_{F}^{G}$
is determined by the gauge group; for the extended model we get from eq.(\ref
{Zs}) and (\ref{solutie}) that 
\begin{eqnarray}
Z_{F}^{G}(M_{z}) &=&\prod_{k=1}^{3}\left[ \frac{\alpha _{g}}{\alpha _{k}(\mu
_{g})}\right] ^{-\frac{2C_{k}(F)}{b_{k}^{\prime }}}\left[ \frac{\alpha
_{k}(\mu _{g})}{\alpha _{k}(M_{z})}\right] ^{-\frac{2C_{k}(F)}{b_{k}}} 
\nonumber \\
&=&\prod_{k=1}^{3}\left[ \frac{\alpha _{g}}{\alpha _{k}(\mu _{g})}\right] ^{%
\frac{2C_{k}(F)}{b_{k}}\frac{n}{b_{k}^{\prime }}}\left[ \frac{\alpha _{g}}{%
\alpha _{k}(M_{z})}\right] ^{-\frac{2C_{k}(F)}{b_{k}}}  \label{zs}
\end{eqnarray}
Note that for one-loop running of the coefficients $Z$ (consistent with two
loop approximation for gauge couplings), the decoupling scale of the heavy
states $\psi $ can be considered to be $\mu _{g}$ rather than a physical
mean mass of the multiplet, as the difference is a radiative effect and
therefore represents a higher (two-loop) correction to $Z$'s or three loop
order for the gauge couplings. This justifies the last equation we wrote
above.

Similarly, we have that in MSSM, 
\begin{equation}  \label{zgmssm}
Z^{o G}_{F}(M_z)=\prod_{k=1}^{3} \left[\frac{\alpha_g^o}{\alpha^o_k(M_z)}%
\right]^{-\frac{2 C_k(F)}{b_k}}
\end{equation}
The coefficients $Z_{F}^{Y}$ can also be expressed in a form similar to that
of $Z_{F}^{G}$ (eq. (\ref{zs})), in terms of Yukawa and gauge couplings at
initial and final scale only; to do this one only needs the one-loop running
for Yukawa couplings; from this, any Yukawa coupling can be expressed (see
Appendix) in terms of the derivative of $\ln y_{\tau,b, t}$ and derivative
of $\ln {\alpha_i}$ where the latter is obtained by replacing $\alpha_i$
with $(1/\tilde b_i)d\ln\alpha_i/dt$, consistent with one-loop running for
Yukawa; hence, the integral $\int y_{\nu}dt$ can be performed analytically
and therefore the differential equation of $Z^Y_F$'s (similar to (\ref{Zs})
without the gauge term) can be integrated to give $Z^Y_F$'s in one-loop
approximation.

A final comment is in order here: we see that the powers entering eq.(\ref
{zs}) depend on $n$ as $n/(b_{i}+n)$. This means that in 
the limit of large $n$ the contribution of these terms
to the running of the couplings brings in 
only a small depedence on n. 
We will later see that $\alpha _{g}/\alpha
_{k}(\mu _{g})$ itself is also relatively stable with respect to $n$. Next,
from the running of the gauge couplings eqs. (\ref{integrated}) and (\ref
{integratedalpha}) we also observe that the {\it explicit} $n$ dependence
comes in under same form for all the couplings as $n/(2\pi )\ln (M_{g}/\mu
_{g})$, which means that the {\it relative} behaviour of the gauge couplings
evolution is affected only through $Z$ factors. With $\alpha _{1}(M_{z})$
and $\alpha _{2}(M_{z})$ fixed to their experimental values, and the above
observations we conclude that the predictions for the strong coupling at
electroweak scale will be stable for large $n$. This observation applies to
other predictions we will make as well, such as the unification scale and
the bare mass, $\mu _{g}$. For the absence of Yukawa couplings' effects this
behaviour was shown in \cite{grl}. We will see that this remains true in
their presence too, although a dependence of the particular type of
superpotential might be expected.

\subsection{Analysis for 3rd generation Yukawa couplings only.}

We will consider first the case when the superpotential is the same in both
MSSM and in the extended model. In eqs. (\ref{mat1}), (\ref{mat2}), (\ref
{mat3}), we replace the gauge part of $Z$ coefficients, $Z_{F}^{G}$, with
their one-loop analytical expressions, consistent with two-loop running for
the gauge couplings. For the Yukawa part of $Z$ coefficients, $Z_{F}^{Y}$,
from their evolution equation 
\begin{equation}  \label{matrixA}
\frac{d}{dt}\ln Z_{F}^{Y}(t)=\sum_{\nu =\tau ,b,t}^{{}}{\cal A}_{\nu }({F}%
)y_{\nu }(t)+{\cal O}(\alpha ^{2})
\end{equation}
after using the relations among the coefficients ${\cal A}_{\nu }(F)$, (not
affected by the presence of the extra matter\footnote{%
Their expressions are presented in Appendix}), one can get the following
relations, valid at any scale between $M_g$ and $M_z$ 
\begin{eqnarray}
Z_{Q_{3}}^{Y} &=&\left( Z_{b_{R}}^{Y}Z_{t_{R}}^{Y}\right) ^{1/2}  \nonumber
\label{wfr} \\
Z_{L_{3}}^{Y} &=&{Z_{\tau _{R}}^{Y}}^{1/2}  \nonumber \\
Z_{H_{1}}^{Y} &=&{Z_{b_{R}}^{Y}}^{3/2}{Z_{\tau _{R}}^{Y}}^{1/2}  \nonumber \\
Z_{H_{2}}^{Y} &=&{Z_{t_{R}}^{Y}}^{3/2}
\end{eqnarray}
where we used, without any loss of generality, the convention $Z^Y(0)=1$ for
any field. Making use of these relations, we get from eq.(\ref{mat1}) that 
\begin{eqnarray}
\delta \alpha _{3}^{-1}(M_{z}) &=&-\frac{470}{77\pi }\ln \left[ \frac{\alpha
_{g}}{\alpha _{g}^{o}}\right] +\sum_{j=1}^{3}\ln \left\{ 1+\frac{%
b_{j}^{\prime }}{n}\left[ \frac{\alpha _{g}}{\alpha _{g}^{o}}-1\right]
\right\} ^{\omega _{j}}  \nonumber  \label{res4} \\
&&+\frac{3}{28\pi }\left[ 5\ln \frac{Z_{b_{R}}^{Y}}{Z_{b_{R}}^{oY}}+\ln 
\frac{Z_{\tau _{R}}^{Y}}{Z_{\tau _{R}}^{oY}}+3\ln \frac{Z_{t_{R}}^{Y}}{%
Z_{t_{R}}^{oY}}\right]
\end{eqnarray}
where $Z^Y$, $\,Z^{oY}$ factors are evaluated at $M_z$ scale and where we
defined 
\begin{equation}\label{omegaaa}
\omega _{j}=\left\{ \frac{-2}{11\pi }\frac{n}{b_{1}^{\prime }},\frac{81}{%
14\pi }\frac{n}{b_{2}^{\prime }},\frac{2}{7\pi }\frac{n}{b_{3}^{\prime }}%
\right\} _{j}
\end{equation}
and $\delta \alpha _{3}^{-1}(M_{z})=\alpha _{3}^{-1}(M_{z})-\alpha
_{3}^{o-1}(M_{z})$. To get the above formula we replaced the terms of the
form $\ln \left( \alpha _{g}/\alpha _{j}(\mu _{g})\right) $ by 
\begin{equation}
\ln \left[ \frac{\alpha _{g}}{\alpha _{j}(\mu _{g})}\right] =\ln \left[
1+\alpha _{g}\frac{b_{j}^{\prime }}{2\pi }\ln \left( \frac{M_{g}}{\mu _{g}}%
\right) \right] =\ln \left[ 1+\frac{b_{j}^{\prime }}{n}\left( \frac{\alpha
_{g}}{\alpha _{g}^{o}}-1\right) \right]
\end{equation}
where we made use of the fact that $\alpha _{g}^{-1}-\alpha
_{g}^{o-1}+n/(2\pi )\ln (M_{g}/\mu _{g})=0$ in one-loop order \footnote{%
In fact $\alpha _{g}^{-1}-\alpha _{g}^{o-1}+b_{i}/(2\pi )\ln \left(
M_{g}/M_{g}^{o}\right) +n/(2\pi )\ln \left( M_{g}/\mu _{g}\right) =0$ in one
loop and we further have $\ln \left( M_{g}/M_{g}^{o}\right) =0$ in one-loop
because the change of unification scale is a two-loop effect.} which is
obtained from equations (\ref{dif12}) and (\ref{al3}) after \footnote{%
We also have that $\delta \alpha _{3}^{-1}$ is non-zero in two-loop only,
being $0$ in one-loop.} noticing that, in one-loop approximation, all
wavefunction renormalisation coefficients are equal to 1.

Similar results hold for the unification scale and for the decoupling scale
of the extra matter. We have from eqs.(\ref{mat2}), (\ref{wfr}) 
\begin{equation}
\frac{M_{g}}{M_{g}^{o}}=\left[ \frac{\alpha _{g}}{\alpha _{g}^{o}}\right] ^{%
\frac{31}{21}}\prod_{j=1}^{3}\left\{ 1+\frac{b_{j}^{\prime }}{n}\left[ \frac{%
\alpha _{g}}{\alpha _{g}^{o}}-1\right] \right\} ^{\rho _{j}}\left[ \frac{%
Z_{b_{R}}^{oY}}{Z_{b_{R}}^{Y}}\right] ^{\frac{1}{7}}\left[ \frac{Z_{\tau
_{R}}^{Y}}{Z_{\tau _{R}}^{oY}}\right] ^{\frac{1}{14}}\left[ \frac{%
Z_{t_{R}}^{oY}}{Z_{t_{R}}^{Y}}\right] ^{\frac{1}{28}}  \label{res1}
\end{equation}
where $Z^Y$, $\,Z^{oY}$ factors are evaluated at $M_z$ scale and 
\begin{equation}
\rho _{j}=\left\{ \frac{1}{12}\frac{n}{b_{1}^{\prime }},{\frac{-39}{28}}{%
\frac{n}{b_{2}^{\prime }}},{\frac{4}{21}}{\frac{n}{b_{3}^{\prime }}}\right\}
_{j}
\end{equation}
Finally from eq.(\ref{mat3}) and (\ref{wfr}) 
\begin{equation}
\frac{\mu _{g}}{M_{g}^{o}}=\left[ \frac{\alpha _{g}}{\alpha _{g}^{o}}\right]
^{\frac{31}{21}+\frac{2336}{231n}}\exp \left[ \frac{2\pi }{n}\left( \alpha
_{g}^{-1}-\alpha _{g}^{o-1}\right) \right] \prod_{j=1}^{3}\left\{ 1+\frac{%
b_{j}^{\prime }}{n}\left[ \frac{\alpha _{g}}{\alpha _{g}^{o}}-1\right]
\right\} ^{\sigma _{j}}\left[ \frac{Z_{\tau _{R}}^{Y}}{Z_{\tau _{R}}^{oY}}%
\right] ^{r_{1}}\left[ \frac{Z_{b_{R}}^{Y}}{Z_{b_{R}}^{oY}}\right] ^{r_{2}}%
\left[ \frac{Z_{t_{R}}^{Y}}{Z_{t_{R}}^{oY}}\right] ^{r_{3}}  \label{res2}
\end{equation}
with 
\begin{equation}
r_{1}={\frac{1}{14}-\frac{3}{7n}},\;\;\;r_{2}={\frac{-1}{7}- \frac{23}{14n}}%
,\;\;\;r_{3}={\frac{-1}{28}-\frac{43}{28n}}
\end{equation}
and 
\begin{equation}
\sigma _{j}=\left\{ {\frac{11n-7}{132\,b_{1}^{\prime }}}, {\frac{-3(111+13n)%
}{28\,b_{2}^{\prime }}}, {\frac{4(22+n)}{21\,b_{3}^{\prime }}}\right\}_{j}
\end{equation}
In the above expressions eqs.({\ref{res4}), ({\ref{res1}), ({\ref{res2}), $%
Z^{Y}$ factors are evaluated at the electroweak scale, and normalised to
unity at the unification scale $M_{g}$ and the same is true for the $Z^{o}$
coefficients\footnote{%
These are normalised to unity at $M_{g}^{o}$!}. If we set all ``Yukawa''
wavefunction coefficients $Z_{F}^{Y}$ and $Z_{F}^{oY}$ equal to unity, we
recover the results \cite{grl} we obtained previously in the absence of any
Yukawa couplings\footnote{%
In ref. \cite{grl}  the results of the extended model
were compared with those of MSSM {\it without} top, bottom, 
tau Yukawa effects.}}}}. As the coefficients $%
Z_{F}^{Y}\geq 1$ (unlike $Z_{F}^{G}\leq 1$) (see eqs.(\ref{Zs}), (\ref
{matrixA})), we note that their effect is to decrease the strong coupling at
electroweak scale. However, we must evaluate the {\it relative} effect of
these coefficients $Z^{Y}$ to those of MSSM, $Z^{oY} $; this will be done
below, by expressing them in terms of Yukawa couplings, assuming the same 
{\it low energy input} for Yukawa couplings in both EMSSM and MSSM models.
This means we take same values at the electroweak scale for top, bottom and $%
\tau $ Yukawa couplings in both models. Using the expressions of $Z^{Y}$ and 
$Z^{oY}$, the result will be expressed as a function of ratios of the Yukawa
couplings, following the details presented at the end of the previous
subsection, after eq.(\ref{zgmssm}).

\noindent For the strong coupling eq.(\ref{res4}) we obtain the following result: $%
1/\alpha _{3}(M_{z})=1/\alpha _{3}^{o}(M_{z})+\delta \alpha _{3}^{-1}(M_{z})$
with 
\begin{eqnarray}  \label{deltaalphastrong}
\delta \alpha _{3}^{-1}(M_{z}) &=&-\frac{11563}{2013\pi }\ln \left[ \frac{%
\alpha _{g}}{\alpha _{g}^{o}}\right] +\sum_{j=1}^{3}\ln \left\{ 1+\frac{%
b_{j}^{\prime }}{n}\left[ \frac{\alpha _{g}}{\alpha _{g}^{o}}-1\right]
\right\} ^{{\cal U}_{j}}  \nonumber \\
&&+\frac{3}{854\pi }\left\{ 4\ln \left[ \frac{y_{\tau }(0)}{y_{\tau }(M_{z})}%
\right] +45\ln \left[ \frac{y_{b}(0)}{y_{b}(M_{z})}\right] +23\ln \left[ 
\frac{y_{t}(0)}{y_{t}(M_{z})}\right] \right\}  \nonumber \\
&&-\frac{3}{854\pi }\left\{ 4\ln \left[ \frac{y_{\tau }^{o}(0)}{y_{\tau
}^{o}(M_{z})}\right] +45\ln \left[ \frac{y_{b}^{o}(0)}{y_{b}^{o}(M_{z})}%
\right] +23\ln \left[ \frac{y_{t}^{o}(0)}{y_{t}^{o}(M_{z})}\right] \right\}
\end{eqnarray}
and 
\begin{equation}
{\cal U}_{j}=\left\{ \frac{-2923}{14091\pi }\frac{n}{b_{1}^{\prime }},\frac{%
4293}{854\pi }\frac{n}{b_{2}^{\prime }},\frac{130}{183\pi }\frac{n}{%
b_{3}^{\prime }}\right\} _{j}
\end{equation}
$j=\{1,2,3\}$. The terms which involve $\alpha_g/\alpha_g^o$ in the r.h.s.
of eq.(\ref{deltaalphastrong}) drive $\alpha_3(M_z)$ above MSSM value. We
also see that, imposing the same electroweak scale values for Yukawa
couplings in both models, the overall effect of the Yukawa couplings depends
on $\ln ({y_{\tau }(0)}/{y_{\tau }^{o}(0)})$, $\ln ({y_{b}(0)}/{y_{b}^{o}(0)}%
)$ and $\ln ({y_{t}(0)}/{y_{t}^{o}(0)})$. The contribution of these terms is
negative and hence $\alpha _{3}(M_{z})$ is further increased from its MSSM
value. The reason for this is that the running of Yukawa couplings is
similar in both models \footnote{%
i.e.they have the same anomalous dimension coefficients}, but in the
extended model they decrease faster due to the fact the gauge couplings in
EMSSM at any scale are increased from their corresponding MSSM values. This
means that, for a fixed low energy input for Yukawa couplings and $SU(2)$
and $U(1)$ gauge couplings, the net effect due the presence of the extra
multiplets is to {\it decrease} the Yukawa couplings at unification scale
while increasing the value of the unified coupling. Hence $y(0)<y^{o}(0)$
for top, bottom and tau quarks and $\alpha_3(M_z)$ is increased. Note that
this is true in the case we have the same superpotential in both models
which means that the coefficients in front of Yukawa couplings in their one
loop running equations are the same to those of MSSM.

The Yukawa effects come in eq.(\ref{deltaalphastrong}) as log's of ratios $%
y_\lambda(0)/y_\lambda(Q)$ with positive sign in front; this might no longer
be true if the superpotential of the model we consider for the extended
model is changed. Still, it seems that a domain with large initial Yukawa
couplings $y_\lambda(0)$ would favour a decreasing effect on the strong coupling at
electroweak scale, particularly when they are larger than their electroweak
scale values. This would in turn favour a fast rate of approach towards
infrared fixed points for ratios of Yukawa to any gauge coupling \cite{sun}.
However, getting a large initial Yukawa coupling, with same low energy input
values as in MSSM is not easy, due to, as we mentioned it above, the gauge
couplings competing effect which seems to be difficult to avoid. In cases
with unified coupling larger than in MSSM, as it happens in our model with
and due to the extra heavy states, the gauge couplings increase while
increasing the scale, causing the Yukawa couplings to decrease as we
increase the scale. This could eventually be compensated for by increasing
the values of the coefficients in front of Yukawa couplings in the one-loop
running of Yukawa couplings. This is possible if the Yukawa structure is
richer than in MSSM; such a particular case will be explored in the next
subsection.

The unification scale is given by (see eq.(\ref{res1})) 
\begin{equation}
\frac{M_{g}}{M_{g}^{o}}=\left[ \frac{\alpha _{g}}{\alpha _{g}^{o}}\right] ^{%
\frac{1711}{1098}}\prod_{j=1}^{3}\left\{ 1+\frac{b_{j}^{\prime }}{n}\left[ 
\frac{\alpha _{g}}{\alpha _{g}^{o}}-1\right] \right\} ^{f_{j}}\prod_{\lambda
=\tau ,b,t}\left[ \frac{y_{\lambda }(0)}{y_{\lambda }(M_{z})}\frac{%
y_{\lambda }^{o}(M_{z})}{y_{\lambda }^{o}(0)}\right] ^{g_{\lambda }}
\label{res1y}
\end{equation}
where 
\begin{equation}
f_{j}=\left\{ \frac{1133}{15372}\frac{n}{b_{1}^{\prime }},\frac{-2277}{1708}%
\frac{n}{b_{2}^{\prime }},\frac{32}{549}\frac{n}{b_{3}^{\prime }}\right\}
_{j}
\end{equation}
and 
\begin{equation}
g_{\lambda }=\left\{ \frac{93}{1708},\frac{-128}{1708},\frac{1}{1708}%
\right\} _{\lambda }  \label{mg}
\end{equation}
with $\lambda =\{\tau,b,t\}$ in this order. The presence of the factor 
\begin{equation}
K=\prod_{\lambda =\tau ,b,t}\left[ \frac{y_{\lambda }(0)}{y_{\lambda }(M_{z})%
}\frac{y_{\lambda }^{o}(M_{z})}{y_{\lambda }^{o}(0)}\right] ^{g_{\lambda }}
\end{equation}
in the expression for $M_{g}/M_{g}^{o}$ prevents us from simplifying any
further this analytical expression. To make predictions one needs, just as
in the previous case, the initial value (at unification scale) of the Yukawa
couplings and their value at the electroweak scale. Since the latter
correspond to the masses of the third family it is reasonable to take them
to be the same in the MSSM and in the extended model. 
Hence $K$ depends on
$y_{\lambda }(0)/y_{\lambda }^{o}(0)$ only. This dependence is very mild
however because the powers $g_{\lambda }$ are very small and therefore, with
a good approximation we have that 
\begin{equation}
K\approx 1
\end{equation}
We checked this numerically\footnote{%
For generic values for Yukawa couplings for high and low $\tan\beta$ case
see ref. \cite{kazakov}.} by using the one-loop running for Yukawa couplings
with one-loop running for the gauge couplings for MSSM case and for the
extended model (in the presence of the extra-matter) and observed that the
ratios $y_{\lambda }^{o}(0)/y_{\lambda }(0)$ are of order 10 giving $K=1-2$,
with larger $K$ (up to 1.8) for larger unified coupling. The main
contribution to increasing the value of $K$ comes from the bottom quark as $%
K $ contains a negative power of $y_b(0)/y^o_b(0)\leq 1$, giving an increase
factor larger than unity, but still very small due to small $g_\lambda$, as
mentioned (this is further suppressed by the tau' contribution). For $%
n=1,2,3 $ case the same discussion about the validity of these results as
that presented in \cite{grl} applies. Hence we have a clear prediction for
the unification scale, 
\begin{equation}
\frac{M_{g}}{M_{g}^{o}}\approx \left[ \frac{\alpha _{g}}{\alpha _{g}^{o}}%
\right] ^{\frac{1711}{1098}}\prod_{j=1}^{3}\left\{ 1+\frac{b_{j}^{\prime }}{n%
}\left[ \frac{\alpha _{g}}{\alpha _{g}^{o}}-1\right] \right\} ^{f_{j}}
\end{equation}
This increase of the unification scale is very close to that found in
reference \cite{grl} and, for $\alpha_g\leq 10 \alpha_g^o\approx 0.4$ is
less than $\approx 3.5$, and depends on the values of $n$, with largest
value for smallest $n$. For large $n$ the results are stable, as we
previously mentioned. We conclude that the effects of Yukawa couplings on
the scale are small and thus there is no need for a further numerical
calculation.

Finally, the bare mass of heavy states is given by 
\begin{equation}
\frac{\mu _{g}}{M_{g}^{o}}=\left[ \frac{\alpha _{g}}{\alpha _{g}^{o}}\right]
^{\frac{1711}{1098}+\frac{104039}{12078n}}\exp {\left[ \frac{2\pi }{n}\left( 
\frac{1}{\alpha _{g}}-\frac{1}{\alpha _{g}^{o}}\right) \right] }%
\prod_{j=1}^{3}\left\{ 1+\frac{b_{j}^{\prime }}{n}\left[ \frac{\alpha _{g}}{%
\alpha _{g}^{o}}-1\right] \right\} ^{{\cal R}_{j}}\prod_{\lambda =\tau ,b,t}%
\left[ \frac{y_{\lambda }(0)}{y_{\lambda }(M_{z})}\frac{y_{\lambda
}^{o}(M_{z})}{y_{\lambda }^{o}(0)}\right] ^{{\cal S}_{\lambda }}  \nonumber
\label{res3}
\end{equation}
\begin{equation}
{\cal R}_{j}=f_{j}+\left\{ \frac{10909}{169092b_{1}^{\prime }},\frac{-15339}{%
1708b_{2}^{\prime }},\frac{1460}{549b_{3}^{\prime }}\right\} _{j}
\end{equation}
\begin{equation}
{\cal S}_{\lambda }=g_{\lambda }+\left\{ \frac{-187}{1708n},\frac{-716}{1708n}%
,\frac{-755}{1708n}\right\} _{\lambda }
\end{equation}
with $j=\{1,2,3\}$ and $\lambda =\{\tau ,b,t\}$. We again take Yukawa
couplings at $M_z$ scale be equal in both MSSM and our extended model. The
dependence of this $\mu_g$ scale on the Yukawa couplings $%
y_\lambda(0)/y^o_\lambda(0)$ is again weak for large $n$, and therefore the
results stay close to those of reference \cite{grl}. We would like to note
that our results for unification scale, the strong 
coupling and decoupling scale $%
\mu_g$ were all computed in terms of Yukawa couplings at the unification and
electroweak scale. The values of these Yukawa couplings can be determined
numerically from their one-loop running only, and we do not need any
numerical work for the gauge couplings running, thus simplifying the
calculation for $M_g$,$\,\alpha_3(M_z)$ and $\,\mu_g$.

\subsection{Family symmetric Yukawa couplings.}

The predictions we make for the strong coupling and the 
unification scale depend on the
type of superpotential we assume; this is so because the running of Yukawa
couplings (and therefore their effect on the running of the gauge couplings)
depends on the type of superpotential. We considered in the previous section
a MSSM type of superpotential and its predictions. We consider now a
different superpotential, to show how these predictions change. The pattern
of the results we present could prove helpful in designing models with
better phenomenological predictions.

We thus turn now to a discussion of the implications of a model designed to
give an acceptable structure for all quark and lepton masses including the
light generations \cite{ibanezross}. In this case the new superpotential has
the following form: 
\begin{equation}  \label{fam_sym_pot}
W=%
\sum_{i,j=1}^{3}(h_{ij}^{u}Q_{i}U_{j}H_{2}^{ij}+h_{ij}^{d}Q_{i}
D_{j}H_{1}^{ij} +h_{ij}^{l}L_{i}e_{j}H_{1}^{ij})
\end{equation}
where the structure in the light quark mass matrix is driven via mixing in
the Higgs sector so that the two light Higgs doublets of the MSSM are
mixtures of the Higgs doublets $H_{1}^{ij}\,$ and of $H_{2}^{ij}.$ This form
is able to reproduce the values of the masses of the third generation for
the case the Yukawa couplings are given by their infra-red fixed point
values in terms of the gauge couplings\cite{sun}. We use this as our
starting point and take the fixed point values for the $h_{ij}^{u}.$ This
means they are all equal, independent of $i,j$. Similarly we take $%
h_{ij}^{d} $ and $h_{ij}^{l}$ to be $i,j$ independent and at their fixed
points. With this we find 
\begin{equation}  \label{w1}
\frac{d}{dt}\ln Z_{F}^{Y}(t)=\sum_{\nu =\tau ,b,t}^{{}}{\cal B} _{\nu }({F}%
)y_{\nu }(t)+{\cal O}(\alpha ^{2})
\end{equation}
where ${\cal B_{\nu }({F})}$ is presented in the Appendix. From eq.(\ref{w1}%
) we obtain the following relations, valid at any scale ${\cal M}$ larger
than $\mu_g$, ${\cal M}\geq\mu_g$ 
\begin{eqnarray}
Z_{Q_{j}}^Y({\cal M})&=&\left[{Z_{b_{R}}^{Y}({\cal M})} {Z_{t_{R}}^{Y}({\cal %
M})} \right] ^{1/2} \\
{Z_{L_{j}}^{Y}({\cal M})}&=& \left[ {Z_{\tau _{R}}^{Y}({\cal M})} \right]%
^{1/2}  \nonumber \\
{Z_{H_{1}^{ij}}^{Y}({\cal M})} &=& \left[ {Z_{b_{R}}^{Y}({\cal M})} \right]%
^{1/2} \left[ {Z_{\tau _{R}}^{Y}({\cal M})} \right]^{1/6}  \nonumber \\
{Z_{H_{2}^{ij}}^{Y}({\cal M})} &=& \left[ {{Z_{t_{R}}^{Y}}({\cal M})} \right]%
^{1/2}  \nonumber
\end{eqnarray}
For the case $M_Z\leq{\cal M}\leq \mu_g$ the above type of superpotential is
no longer valid and a MSSM-like superpotential applies. This is so because
the extra heavy states we consider, among which are all Higgs fields%
\footnote{%
This implies that the number of extra heavy states is larger than 5} of eq. (%
\ref{fam_sym_pot}) decouple at $\mu_g$. Below this scale $\mu_g$, only the
third generation' Yukawa couplings give important contributions to
wavefunction renormalisation coefficients $Z^Y_{\phi_k}$, while for the
first two generations the factors $Z_{\phi_j}$, $\,j=\{1,2\}$ evolve only
through their gauge contributions, $Z_{\phi_j}^G$,$\,j=\{1,2\}$. We thus
have that for $M_Z\leq{\cal M}\leq \mu_g$ 
\begin{equation}  \label{ww1}
\frac{d}{dt}\ln Z_{F}^{Y}(t)=\sum_{\nu =\tau ,b,t}^{{}}{\cal A} _{\nu }({F}%
)y_{\nu }(t)+{\cal O}(\alpha ^{2})
\end{equation}
which, after integration below $\mu_g$, gives that 
\begin{eqnarray}
\frac{Z_{Q_{3}}^Y({\cal M})}{Z_{Q_{3}}^{Y}(\mu_g)} &=&\left[ \frac{%
Z_{b_{R}}^{Y}({\cal M})}{ Z_{b_{R}}^{Y}(\mu_g)} \frac{Z_{t_{R}}^{Y}({\cal M})%
}{Z_{t_{R}}^{Y}(\mu_g)} \right] ^{1/2} \\
\frac{Z_{L_{3}}^{Y}({\cal M})}{Z_{L_{3}}^{Y}({\mu_g})} &=& \left[ \frac{%
Z_{\tau _{R}}^{Y}({\cal M})}{Z_{\tau _{R}}^{Y}(\mu_g)} \right]^{1/2} 
\nonumber \\
\frac{Z_{H_{1}}^{Y}({\cal M})}{Z_{H_{1}}^{Y}(\mu_g)} &=& \left[ \frac{%
Z_{b_{R}}^{Y}({\cal M})}{Z_{b_{R}}^{Y}({\mu_g})} \right]^{3/2} \left[ \frac{%
Z_{\tau _{R}}^{Y}({\cal M})}{Z_{\tau_{R}}^{Y}(\mu_g)} \right]^{1/2} 
\nonumber \\
\frac{Z_{H_{2}}^{Y}({\cal M})}{Z_{H_{2}}^{Y}({\mu_g})} &=& \left[ \frac{{%
Z_{t_{R}}^{Y}}({\cal M})}{{Z_{t_{R}}^{Y}}({\mu_g})} \right]^{3/2}  \nonumber
\end{eqnarray}
Note also the similarity with eq. (\ref{wfr}), as expected, with the
difference that the initial condition for any $Z^Y$ takes place at $\mu_g$
scale (with $Z^Y(\mu_g)\not=1$!) rather than at unification scale.

Following the procedure outlined above we obtain, in this case, the
following change to the strong coupling: 
\begin{eqnarray}
\delta \alpha _{3}^{-1}(M_{z}) &=&-\frac{470}{77\pi }\ln \left[ \frac{\alpha
_{g}}{\alpha _{g}^{o}}\right] +\sum_{j=1}^{3}\ln \left\{ 1+\frac{%
b_{j}^{\prime }}{n}\left[ \frac{\alpha _{g}}{\alpha _{g}^{o}}-1\right]
\right\} ^{\omega _{j}}  \nonumber  \label{WWW} \\
&&+\frac{9}{28\pi }\left[ \ln {Z_{b_{R}}^{Y}(\mu _{g})}-\ln {\
Z_{t_{R}}^{Y}(\mu_{g})}-\frac{1}{3}\ln {\ Z_{\tau _{R}}^{Y}(\mu _{g})}\right]
\nonumber \\
&&+\frac{3}{28\pi} \left[5\ln\frac{Z_{b_{R}}^{Y}(M_{z})}{Z_{b_{R}}^{Y}({\mu_g%
})}+ \ln \frac{Z_{\tau_{R}}^{Y}(M_{z})}{Z_{\tau_{R}}^{Y}(\mu_g)} +3\ln \frac{%
Z_{t_{R}}^{Y}(M_z)}{Z_{t_{R}}^{Y}(\mu_g)}\right]  \nonumber \\
&&-\frac{3}{28\pi }\left[ 5\ln {Z_{b_{R}}^{oY}(M_{z})}+\ln{Z_{\tau
_{R}}^{oY}(M_{z})}+3\ln {Z_{t_{R}}^{oY}(M_{z})}\right]
\end{eqnarray}
where 
\begin{equation}
\omega _{j}=\left\{ \frac{-2}{11\pi }\frac{n}{b_{1}^{\prime }},\frac{81}{%
14\pi }\frac{n}{b_{2}^{\prime }},\frac{2}{7\pi }\frac{n}{b_{3}^{\prime }}%
\right\} _{j}
\end{equation}
and $\delta \alpha _{3}^{-1}(M_{z})=\alpha _{3}^{-1}(M_{z})-\alpha
_{3}^{o-1}(M_{z})$. Setting all $Z^Y$ equal to unity in (\ref{WWW}) gives
the previous result \cite{grl} where only gauge effects on $\alpha_3(M_z)$
were considered, leading to an increase of its value from the MSSM value. We
also see that the Yukawa effects give in this case two negative signs in eq.(%
\ref{WWW}) in the square bracket which make its contribution rather small
and negative; For comparison see the result of eq.(\ref{res4}) for the MSSM
superpotential. This contribution, if the signs were positive was expected
to be the largest as $Z$ factors have a steeper running between $M_g$ (where
they are equal to unity) and $\mu_g$ than between $\mu_g$ and $M_z$ due to
larger coefficients in (\ref{w1}) than in (\ref{ww1}). The second-last and
last square brackets in eq.(\ref{WWW}) give a small correction to alpha
strong and in fact they largely cancel each other. We can further express $%
\delta\alpha_3^{-1}(M_z)$ in function of Yukawa couplings evaluated at $M_g$%
, at $\mu_g$ and at electroweak scale. 
\begin{eqnarray}  \label{deltaalpha}
\delta\alpha^{-1}_3(M_z)&=& -\frac{11563}{2013\pi}\ln\left[\frac{\alpha_g}{%
\alpha_g^o}\right] +\sum_{i=1}^{3}{\cal Z}_j\ln \left\{ 1+\frac{b_{j}^{\prime
}}{n} \left[\frac{\alpha_g}{\alpha_{g}^{o}}-1\right]\right\}  \nonumber \\
&&+\left\{\frac{243}{1022\pi}\ln\left[\frac{y_b(0)}{y_b(\mu_g)}\right] -%
\frac{45}{511\pi}\ln\left[\frac{y_{\tau}(0)}{y_{\tau}(\mu_g)}\right] -\frac{%
225}{1022\pi}\ln\left[\frac{y_t(0)}{y_t(\mu_g)}\right]\right\}  \nonumber \\
&&+\left\{\frac{135}{854\pi}\ln \left[\frac{y_b(\mu_g)}{y_b(M_z)}\right] +%
\frac{6}{427\pi}\ln\left[\frac{y_\tau(\mu_g)}{y_\tau(M_z)}\right] +\frac{69}{%
854\pi}\ln\left[\frac{y_t(\mu_g)}{y_t(M_z)}\right]\right\}  \nonumber \\
&&-\left\{\frac{135}{854\pi}\ln \left[\frac{y_b^o(0)}{y_b^o(M_z)}\right] +%
\frac{6}{427\pi}\ln\left[\frac{y_\tau^o(0)}{y_\tau^o(M_z)}\right] +\frac{69}{%
854\pi}\ln\left[\frac{y_t^o(0)}{y_t^o(M_z)}\right]\right\}
\end{eqnarray}
with ${\cal Z}_j$ given by 
\begin{equation}
{\cal Z}_j= \left\{\frac{1}{\pi}\left(\frac{-2923}{14091}+\frac{351}{365
b_1^{\prime}}\right), \frac{1}{\pi}\left(\frac{4293}{854}+\frac{-6129}{1022
b_2^{\prime}}\right), \frac{1}{\pi}\left(\frac{130}{183}+\frac{486}{511
b_3^{\prime}}\right)\right\}_j
\end{equation}
The terms in (\ref{deltaalpha}) which contain $\alpha_g/\alpha_g^o$ have an
overall increasing effect on the strong coupling 
driving it towards values larger
than in MSSM as it can be seen by simply plotting their sum with respect to $%
\alpha_g$. 
We must therefore analyse the effect Yukawa
couplings in eq.(\ref{deltaalpha}) have. Their effect on $\alpha_3(M_z)$ is
not always opposite to that of the gauge couplings. In the expression above
we see three types of contributions. The first one is due to Yukawa effects
between unification scale and (effective) decoupling scale $\mu_g$. This is
due to the family symmetric Yukawa couplings we chose above this scale. We
can {\it assume} that the values of $y_\lambda(0)$ are larger than $%
y_\lambda(\mu_g)$; this would be expected given that we considered the
family symmetric couplings, which is possible if they flow to infrared fixed
point values, thing favoured by a large initial Yukawa\footnote{%
Considering $y_\lambda(0)$ smaller than $y_\lambda(\mu_g)$ is not allowed in
this model because a rapid flow to infrared fixed points was assumed here
and hence generation independent Yukawa couplings are possible for $%
y_\lambda(Q)/y_\lambda(0)\leq 1$, with $Q\leq M_g$}. The overall effect
would therefore be to increase the strong 
coupling, due to the negative signs of
the last two terms for tau and top couplings in the first curly bracket eq.(%
\ref{deltaalpha}) which dominate the bottom contribution. This is unlike
equation (\ref{deltaalphastrong}), where the log's of Yukawa ratios had
positive sign in front, (but were themselves negative). Here they are
positive, but their sign in front is not positive for all bottom, tau, top
couplings.

The second contribution comes from terms like $y_\lambda(\mu_g)/y_%
\lambda(M_z)$ and the third contribution comes from MSSM terms, $%
y^o_\lambda(0)/y^o_\lambda(M_z)$. These terms largely cancel because the
running of Yukawa couplings below $\mu_g$ scale is the same in MSSM and
EMSSM and we have same electroweak values for $y_\lambda(M_z)$, $%
y^o_\lambda(M_z)$ (these are inputs of our analysis); We also know that
Yukawa effects in MSSM, coming in as ratios $y^o_\lambda(0)/y^o_\lambda(M_z)$
are small and the scale $\mu_g$ being in general heavy\footnote{%
The extra states are not protected by any chiral symmetry and therefore
their mass is heavy}, we conclude that the last two curly brackets of eq.(%
\ref{deltaalpha}) do not solve the problem of how to reduce the small
discrepancy between the predicted value for the strong coupling and its
experimental value.

To conclude, it is not easy to reduce the strong coupling and increase the
unification scale significantly, and it seems that the observation we made
after eq.(\ref{mat2}) is true in general for models which consider only {\it %
complete} additional representations. 

\section{The large $n$ limit.}

Our analysis has so far assumed that perturbation theory works well up to
the unification scale, and that the presence of a given number of states
does not affect its convergence. However, for a large number of states {\it %
and} a large unified coupling one should carefully consider the limits of
the two-loop perturbative expansion we applied in this work as well as in
ref. \cite{grl}. It is the purpose of this section to explore the
phenomenological implications for the case when these limits are reached, in
the absence of Yukawa effects.

In the case of large $n$ and large $\alpha _{g}$ the perturbative expansion
breaks down. An estimate of where breakdown occurs can be obtained by
comparing two-loop beta function terms with three loop terms\footnote{%
For this, expand the denominator of eq. (\ref{shifmanbetaalpha}) to get
three loop terms} or, equivalently, one-loop terms in the expansion of the
anomalous dimensions of the fields with two-loop terms. Although the higher
order terms have an additional power of coupling they are also proportional
to $n$ and in the large n limit this will compensate for the additional
coupling. This happens for $n\alpha \approx n^{2}\alpha ^{2}/(4\pi )$, or $%
n\alpha \approx {\cal O}(4\pi )$. \ In the case of large n the perturbation
series in $\alpha $ as well as the perturbation series for anomalous
dimensions of chiral fields can be resummed \cite{largeN} to leading order $%
O(1/n)$. This calculation was done by Jones in \cite{largeN} and the reader
is referred to this work for full details. We will use their results for the
resummed anomalous dimensions of the light fields of the spectrum for the
case the Yukawa couplings are ignored.

The anomalous dimension for any matter field is given by 
\begin{equation}
\gamma _{F}=-\sum_{k=1}^{3}\frac{\alpha _{k}}{2\pi }G\left( \frac{\alpha
_{k}n}{4\pi }\right) C_{k}(R_{F})
\end{equation}
Here $F$ stands for a matter field in the representation $R_{F}$ of the
gauge group $SU(k)$\footnote{%
For the U(1) factor $C_{k}(R_{F})$ is $\frac{3}{5}Y^{2}$ where $Y$ is the
usual weak hypercharge \cite{largeN}.}. The function G is defined by 
\begin{equation}
G(\epsilon )=\frac{1}{2}\frac{\Gamma (3-2\epsilon )}{\Gamma (2-\epsilon )^{2}%
}\frac{\sin (\pi \epsilon )}{\pi \epsilon }
\end{equation}
and has a pole for $\epsilon =3/2$ which sets the radius of convergence of
the {\it resummed} series. For $G=1$ we recover the results of eq.(\ref{Zs}%
). The running of the gauge couplings is now given by 
\begin{eqnarray}
\alpha _{i}^{-1}(M_{z})& =-\delta _{i}+\alpha _{g}^{-1}+\frac{b_{i}}{2\pi }%
\ln \left[ \frac{M_{g}}{M_{z}}\right] +\frac{n}{2\pi }\ln \left[ \frac{M_{g}%
}{\mu _{g}}\right] +\frac{1}{4\pi }\sum_{j=1}^{3}\frac{n}{b_{j}^{\prime }}%
\left[ 2\delta _{ij}\lambda _{j}-\frac{b_{ij}}{b_{j}}\right] \ln \left[ 
\frac{\alpha _{g}}{\alpha _{j}(\mu _{g})}\right]  \nonumber \\
&  \nonumber \\
& +\frac{1}{4\pi }\sum_{j=1}^{3}\frac{b_{ij}}{b_{j}}\ln \left[ \frac{\alpha
_{g}}{\alpha _{j}(M_{z})}\right] +{\cal K}_{i}  \label{shifman}
\end{eqnarray}
\begin{table}[tbp]
\begin{center}
\begin{tabular}{|c|c|c|cr|}
\hline
$n$ & $\alpha_g$ & $M_g/M_g^o$ & $\alpha_3(M_z)$ &  \\ \hline\hline
30 & 0.1 & 1.359 & 0.12700 &  \\ \hline
30 & 0.2 & 1.760 & 0.12796 &  \\ \hline
30 & 0.3 & 2.043 & 0.12850 &  \\ \hline
40 & 0.1 & 1.355 & 0.12698 &  \\ \hline
40 & 0.2 & 1.745 & 0.12786 &  \\ \hline
40 & 0.3 & 2.019 & 0.12835 &  \\ \hline
50 & 0.1 & 1.353 & 0.12696 &  \\ \hline
50 & 0.2 & 1.737 & 0.12780 &  \\ \hline
50 & 0.3 & 2.006 & 0.12828 &  \\ \hline
60 & 0.1 & 1.351 & 0.12695 &  \\ \hline
60 & 0.2 & 1.732 & 0.12777 &  \\ \hline
60 & 0.3 & 1.999 & 0.12823 &  \\ \hline
\end{tabular}
\end{center}
\caption{The values of $\protect\alpha _{g}$, strong coupling at $\protect%
\alpha _{3}(M_{Z})$, and $M_{g}/M_{g}^{o}$ as a function of (large) n. We
always have that $(n\protect\alpha _{g})/(4\protect\pi )<3/2$.}
\label{table:1}
\end{table}
where $b_{ij}$ and $\,b_{j}$ are just the two-loop and one-loop beta
functions for the MSSM and where $\lambda_1=0$, $\,\lambda_2=2$,
$\,\lambda_3=3$. ${\cal K}_{i}$ are the resummed corrections induced
in the large $n$ limit ignoring Yukawa effects and are given by 
\begin{equation}
{\cal K}_{i}=\frac{1}{4\pi }\sum_{j=1}^{3}(b_{ij}-2\lambda _{j}\delta
_{ij}b_{j})\frac{1}{b_{j}^{\prime }}\int_{\alpha _{j}(\mu _{g})}^{\alpha
_{g}}\frac{d\alpha _{j}}{\alpha _{j}}\left[ -1+G\left( \frac{\alpha _{j}n}{%
4\pi }\right) \right]  \label{kapa}
\end{equation}
The effect of ${\cal K}_{i}$'s on the unification scale and the strong
coupling may now be readily computed to give 
\begin{equation}
\frac{M_{g}}{M_{g}^{o}}=e^{\frac{5\pi }{14}(-{\cal K}_{1}+{\cal K}_{2})}%
\left[ \frac{\alpha _{g}}{\alpha _{g}^{o}}\right] ^{31/21}\left[ \frac{%
\alpha _{3}(M_{z})}{\alpha _{3}(M_{z})^{o}}\right] ^{4/21}\left[ \frac{%
\alpha _{g}}{\alpha _{1}(\mu _{g})}\right] ^{n/(12b_{1}^{\prime })}\left[ 
\frac{\alpha _{g}}{\alpha _{2}(\mu _{g})}\right] ^{-39n/(28b_{2}^{\prime })}%
\left[ \frac{\alpha _{g}}{\alpha _{3}(\mu _{g})}\right] ^{4n/(21b_{3}^{%
\prime })}  \label{emg}
\end{equation}
and 
\begin{equation}
\alpha _{3}^{-1}(M_{z})=\alpha _{3}^{o-1}(M_{z})-\frac{470}{77\pi }\ln \left[
\frac{\alpha _{g}}{\alpha _{g}^{o}}\right] +\sum_{j=1}^{3}\ln \left[ \frac{%
\alpha _{g}}{\alpha _{j}(\mu _{g})}\right] ^{\omega _{j}}-\frac{17}{14\pi }%
\ln \left[ \frac{\alpha _{3}(M_{z})}{\alpha _{3}^{o}(M_{z})}\right] +\left( 
{\cal K}_{3}-\frac{12}{7}{\cal K}_{2}+\frac{5}{7}{\cal K}_{1}\right)
\label{alst}
\end{equation}
with $\omega_j$ given in eq.(\ref{omegaaa}).
To use these equations it is necessary to compute the values of $\alpha
_{j}(\mu _{g}).$ We do this by writing the equations equivalent to eq.(\ref
{shifman}) for $\alpha _{i}^{-1}(\mu _{g})$ and eliminating the term $n/(2\pi
)\ln \left( M_{g}/\mu _{g}\right) $ between these equations for $\alpha
_{i}. $ This gives 
\begin{equation}
\alpha _{i}^{-1}(\mu _{g})=\alpha _{g}^{-1}+\frac{5}{28}\frac{b_{i}^{\prime }%
}{n}\left( b_{2}{\cal K}_{1}-b_{1}{\cal K}_{2}\right) +{\cal K}_{i}-\frac{%
b_{i}^{\prime }}{n}\left[ \frac{1}{\alpha _{g}}-\frac{1}{\alpha _{g}^{o}}%
\right] -\frac{2336}{231n}\frac{b_{i}^{\prime }}{2\pi }\ln \left[ \frac{%
\alpha _{g}}{\alpha _{g}^{o}}\right] +
\sum_{j=1}^{3}\ln \left[ \frac{\alpha _{g}}{\alpha _{j}(\mu _{g})}\right]
^{v_{ij}}  \label{alphas}
\end{equation}
with 
\begin{equation}
v_{ij}=\frac{1}{4\pi b_{j}^{\prime }}(b_{ij}+2n\lambda _{j}\delta _{ij})+%
\frac{b_{i}^{\prime }}{2\pi }\left\{ \frac{7}{132b_{1}^{\prime }},\frac{333}{%
28b_{2}^{\prime }},\frac{-88}{21b_{3}^{\prime }}\right\} _{j}
\end{equation}
Note that the quantities ${\cal K}_{i}$ depend on $\alpha _{j}(\mu _{g})$ as
seen from eq.(\ref{kapa}). For given $\alpha _{g}$ and large $n$ we solve (%
\ref{alphas}) to get $\alpha _{j}(\mu _{g})$ which are then used to compute
the ratio $M_{g}/M_{g}^{o}$ and $\alpha _{3}(M_{z})$ of eqs. (\ref{emg}), (%
\ref{alst}). The results for values of $n$ and $\alpha _{g}$ which avoid the
pole in $G$ are presented in Table 1.

We see that the effects of a large number of states is very small and the
results for $M_{g}/M_{g}^{o}$ and $\alpha _{3}(M_{z})$ are in general
insensitive to the variation of (large) $n$ and stay close to the MSSM
values.

\section{The case of strong coupling.}

Our analysis to date has assumed that the couplings (even at large n) remain
in the perturbative domain between electroweak scale and unification scale.
However, non-perturbative physics could be important {\it below unification
scale} and in this section we will consider this possibility. In this case
one apparently loses all the predictive power because the full beta
functions are known only perturbatively and the perturbation series does not
work beyond the scale of non-perturbative physics. In \cite{sun} it was
argued that this is not the case because the {\it ratios} of the gauge
couplings are driven towards stable infra-red fixed points. Since the
couplings are initially large, the rate of flow to the infrared fixed point
should be rapid. As a result, the low-energy values of these ratios are
insensitive to their initial values and to the non-perturbative effects
(provided the couplings initially lie in the domain of attraction of the
fixed points), giving a reliable prediction for $\alpha _{3}(M_{z})$ even in
this case. Using the fixed points (FP) of the one-loop RGE to determine the
boundary conditions of the gauge couplings at the decoupling scale of the
additional massive states, it was found that a low value of $\alpha
_{3}(M_{z})$ is predicted, very close to the mean experimental value and in
much better agreement than the perturbative MSSM result.

Here we wish to consider corrections to the above ``fixed point'' result in
order to estimate the precision of the result and to establish whether the
apparent improvement over the perturbative case is significant. To do this
we shall continue to use the perturbative solution, but only in the domain
where it is applicable. As we shall see in this region one may see the RG
flow to the fixed points but there are calculable corrections. The first
point to make is that even using the improved perturbation sum discussed in
the last section one requires\footnote{We will however ask that
$n\alpha \leq {\cal O}(4\pi)$ instead, because we will not 
use the resummed perturbation series, and this relation marks the limit
where it becomes important.} 
$n\alpha \leq {\cal O}(6\pi ).$ We can only
calculate those effects below the non-perturbative domain at the scale $M_{o}
$ where the gauge couplings enter the perturbative domain and as we have
just remarked this starts at quite small values of $\alpha $. Thus our error
estimates will necessarily be somewhat rough. There are two corrections to
the ``fixed point'' calculation, deriving from the relations among the gauge
couplings at the decoupling scale\footnote{%
In two-loop running of gauge couplings the effective decoupling scale is $%
\mu _{g}$, the bare mass of the heavy spectrum} $\mu _{g}$, and from the
uncertainty in their exact values as they leave the non-perturbative region
to enter the perturbative domain. Here we analyse them briefly.

The first correction arises because, even if one sticks to the one-loop beta
functions, due to the finite energy range involved the ratios of couplings
are not driven quite to the fixed point. Therefore the boundary condition we
used for the evolution of the couplings below $\mu _{g}$ is {\it not} the
``fixed-point'' ratio \cite{sun} 
\begin{equation}
R_{ji}(\mu _{g})\equiv \frac{\alpha _{j}(\mu _{g})}{\alpha _{i}(\mu _{g})}%
=R_{ji}^{\ast }  \label{fixedpoint}
\end{equation}
with $R_{ji}^{\ast }=b_{i}^{\prime }/b_{j}^{\prime }$. Instead, the boundary
condition for the running of the gauge couplings below the decoupling scale%
\footnote{%
Below this scale a MSSM like spectrum applies} is a ``quasi-fixed-point''
(QFP) relation for the ratio of the gauge couplings. This relation exactly
takes into account a one-loop running for the gauge couplings above $\mu
_{g} $ scale (just as in FP case) {\it and} the {\it finite} range of
energy. This QFP relation is 
\begin{equation}
R_{ji}(\mu _{g})\equiv \frac{\alpha _{j}(\mu _{g})}{\alpha _{i}(\mu _{g})}=%
\frac{R_{ji}^{\ast }}{1-\left[ 1-\frac{R_{ji}^{\ast }}{R_{ji}(M_{o})}\right] 
\frac{\alpha _{i}(\mu _{g})}{\alpha _{i}(M_{o})}}  \label{qfp}
\end{equation}
This can be easily deduced by integrating the one-loop differential eqs for
the gauge couplings above $\mu _{g}$. Using this QFP boundary condition for
the gauge couplings evolution below $\mu _{g}$ scale, one gets a value for $%
\alpha _{3}(M_{z})$ which is typically 
smaller than that of FP case. The results
depend on the value of $R_{ij}(M_{o})$, which enters (\ref{qfp}), but this
dependence is relatively weak and the prediction for $\alpha _{3}(M_{z})$
stays below MSSM value. This is somewhat expected, as we know that two-loop
corrections in MSSM increase $\alpha _{3}(M_{z})$ from its rather good one
loop prediction\footnote{%
In MSSM a one-loop calculation gives $\alpha _{3}(M_{z})\approx 0.117$ while
two-loop calculation increases it to $\approx 0.126$}, taking it away from
the current experimental upper limit. As eq. (\ref{qfp}) ignores the two
loop evolution' effects between $\mu _{g}$ and $M_{o}$, we get a smaller $%
\alpha _{3}(M_{z})$ than in MSSM.

This brings us to the second correction which consists of two-loop or higher
order (gauge and Yukawa) effects between $\mu _{g}$ and $M_{o}$. This
affects the relation (\ref{qfp}) and therefore the low energy predictions
for the strong coupling. We will consider 
here only the gauge effects. Their effect
is to change the expression of $R_{ji}(\mu _{g})$ such as to increase the
value of the strong coupling at electroweak scale. 
The increase overcomes the QFP
decrease in $\alpha _{3}(M_{z})$, bringing it closer to MSSM prediction.
This conclusion is subject to the corrections due to the unknown values of
the gauge couplings at $M_{o}$ (or equivalently $Z(M_{o})$), where they
enter the perturbative domain as we lower the scale, but there is a
systematic effect increasing the value of $\alpha _{3}(M_{z})$ even for
sizeable changes in the boundary conditions.

Our analysis is limited because the range between $\mu _{g}$ and $M_{o}$ is
quite small and the {\it calculable} corrections to the fixed point result
correspondingly small. However this does not mean the full non-perturbative
correction to the fixed point result is small. An estimate of these
corrections coming from the higher order and quasi fixed point terms
discussed above may be obtained by noting that there is a focussing effect
of the fixed point following in the perturbative flow domain which reduces
the contribution of such effects by the factor $\frac{\alpha _{i}(\mu _{g})}{%
\alpha _{i}(M_{o})}$ (cf eq(\ref{qfp})). Using this factor with  $\alpha
_{i}(M_{o})$ limited by the perturbative condition $n\alpha \leq {\cal O}%
(4\pi )$ together with the assumption of $O(1)$ deviations from the QFP
boundary condition we arrive at a reasonably conservative estimate for these
errors of ${\cal O}(\pm 0.01)$ for $n=12$, increasing for larger
values of $n$.

\section{Summary and Conclusions}

For the case of perturbative evolution of the gauge and Yukawa couplings up
to the unification scale, we derived an analytical method to determine the
two-loop unification predictions for the value of unification scale, the
intermediate mass scale and the value of the 
strong coupling at $M_{z}$ in models
with the MSSM spectrum augmented by additional massive representations
filling out complete $SU(5)$ representations. The effects of the two-loop
terms involving Yukawa couplings are in general relatively small and
model-dependent. For the models we considered, $\alpha _{3}(M_{z})$ cannot
be lowered below the MSSM value, even with Yukawa effects present, keeping $%
\alpha _{3}(M_{z})$ above the mean experimental value. However the sign of
the effect is not universal, so it is possible a richer structure in Yukawa
sector could reduce the strong coupling. 
We also showed that the unification scale
is not changed significantly by the top, bottom and tau Yukawa coupling
effects. Using the large n resummed perturbation series we showed that the
results are stable in the limit of inclusion of a large number of complete $%
SU(5)$ representations and stay close to MSSM predictions. Finally we
considered the case that unification occurs at strong coupling with
perturbative analysis breaking down {\it below} the unification scale.
Because of the fixed point structure in the RGEs (at one-loop level) one may
still make a  prediction for $\alpha _{3}(M_{z})$. However threshold and two
loop effects above the decoupling scale for the heavy states can be
sizeable. As a result the low estimate for $\alpha _{3}(M_{z})$ obtained
using fixed point boundary conditions at the decoupling scale is increased
for large n towards the MSSM value. Overall we estimate an irreducible error
in the strong coupling determination of $\alpha _{3}(M_{z})$ of $O(0.01)$
coming from the residual sensitivity to the initial values of the ratios of
the couplings as they enter the perturbative domain and from two-loop
corrections to $R_{ji}(\mu _{g})$ above $\mu _{g}$.

\section{Acknowledgments}

D.G. gratefully acknowledges the financial support from the part of
University of Oxford and Oriel College (University of Oxford). G. A.-C.
acknowledges the financial support for his work on this project from
P.P.A.R.C., the Foundation Blanceflor Boncompagni-Ludovisi and the Swiss
National Science Foundation.

\section{Appendix}

We have the following expressions for $T(R_\sigma)$ for the fundamental
representation 
\begin{equation}
\delta^{ab}T(R)=Tr(T^aT^b)=\frac{1}{2}\delta^{ab}
\end{equation}
For a given flavour the value of $T(R_\sigma)$ is $T(R_\sigma)=1/2$. ($%
T(R_\sigma)=1$ for conjugate pairs of fields).\newline
\noindent For the adjoint representation we have 
\begin{equation}
\left(T^a\right)_{bc}=-if^{abc}
\end{equation}
and therefore, for $SU(N)$ group 
\begin{equation}
\delta^{ab}T(G)=f^{acd}f^{bcd}=N\delta^{ab}
\end{equation}
where the structure constants $f^{abc}$ are given by 
\begin{equation}
\left[T^a,T^b\right]=if^{abc} T^c
\end{equation}
The values of ${\cal A}$ used in text (\ref{matrixA}) in the context of a
MSSM like superpotential and considering only the Yukawa couplings for top,
bottom and tau is given by (see for example \cite{bjorkman_jones}) 
\begin{equation}
{\cal A}_{\nu}(F)=\left( 
\begin{array}{crcrcrcrcrcrcrcr}
& l_{L_3} & q_{L_3} & e_{R_3} & u_{R_3} & d_{R_3} & H_1 & H_2 \\
\\
\nu=\tau:\; & -1 & 0 & -2 & 0 & 0 & -1 & 0  \\ 
\nu=b: \; & 0 & -1 & 0 & 0 & -2 & -3 & 0     \\ 
\nu=t: \; & 0 & -1 & 0 & -2 & 0 & 0 & -3 
\end{array}
\right)
\end{equation}
The running of Yukawa couplings in one-loop, for a MSSM like superpotential
can be rewritten, using \cite{bjorkman_jones}, as follows (we ignore the
first two generations' contribution) 
\begin{equation}
\left( 
\begin{array}{c}
y_\tau(t) \\ 
\\ 
y_b(t) \\ 
\\ 
y_t(t)
\end{array}
\right) =\left( 
\begin{array}{crcrcr}
\frac{35}{122} & \frac{-9}{61} & \frac{3}{122} \\
\\
\frac{-3}{61} & \frac{12}{61} & \frac{-2}{61}  \\
\\
\frac{1}{122} & \frac{-2}{61} & \frac{21}{122} 
\end{array}
\right) \left( 
\begin{array}{c}
\frac{d \ln y_\tau}{dt} \\ 
\\ 
\frac{d \ln y_b}{dt} \\ 
\\ 
\frac{d \ln y_t}{dt}
\end{array}
\right) +\left( 
\begin{array}{crcrcr}
\frac{143}{305} & \frac{30}{61} & \frac{-40}{61} \\
\\
\frac{-23}{915} & \frac{21}{61} & \frac{160}{183} \\
\\
\frac{136}{915} & \frac{27}{61} & \frac{136}{183} \\ 
\end{array}
\right) \left( 
\begin{array}{c}
\alpha_1(t) \\ 
\\ 
\alpha_2(t) \\ 
\\ 
\alpha_3(t)
\end{array}
\right)+{\cal O}(\alpha^2)
\end{equation}
For a family symmetric superpotential we used in text (\ref{fam_sym_pot}) 
\begin{equation}
W=\sum_{i,j=1}^{3}(h_{ij}^{u}Q_{i}U_{j}H_{2}^{ij}
+h_{ij}^{d}Q_{i}U_{j}H_{1}^{ij}+h_{ij}^{l}L_{i}e_{j}H_{1}^{ij})
\end{equation}
we have that 
\begin{equation}
{\cal B}_{\nu}(F)=\left( 
\begin{array}{crcrcrcrcrcrcrcr}
& l_{L_j} & q_{L_j} & e_{R_j} & u_{R_j} & d_{R_j} & H_1 & H_2 \\
\nu=\tau:\; & -3 & 0 & -6 & 0 & 0 & -1 & 0   \\ 
\nu=b: \; & 0 & -3 & 0 & 0 & -6 & -3 & 0   \\ 
\nu=t: \; & 0 & -3 & 0 & -6 & 0 & 0 & -3  
\end{array}
\right)
\end{equation}
For the above class of superpotential, the Yukawa couplings running in one
loop can be written as follows 
\begin{equation}
\left( 
\begin{array}{c}
y_\tau(t) \\ 
\\ 
y_b(t) \\ 
\\ 
y_t(t)
\end{array}
\right) =\left( 
\begin{array}{crcrcr}
\frac{15}{146} & \frac{-2}{73} & \frac{1}{146} \\
\\
\frac{-2}{219} & \frac{20}{219} & \frac{-5}{219} \\
\\
\frac{1}{438} & \frac{-5}{219} & \frac{13}{146} \\
\end{array}
\right) \left( 
\begin{array}{c}
\frac{d \ln y_\tau}{dt} \\ 
\\ 
\frac{d \ln y_b}{dt} \\ 
\\ 
\frac{d \ln y_t}{dt}
\end{array}
\right) +\left( 
\begin{array}{crcrcr}
\frac{13}{73} & \frac{18}{73} & \frac{-8}{73} \\
\\
\frac{7}{1095} & \frac{13}{73} & \frac{80}{219} \\
\\
\frac{232}{3285} & \frac{15}{73} & \frac{232}{657} \\
\end{array}
\right) \left( 
\begin{array}{c}
\alpha_1(t) \\ 
\\ 
\alpha_2(t) \\ 
\\ 
\alpha_3(t)
\end{array}
\right)+{\cal O}(\alpha^2)
\end{equation}

\end{document}